\documentclass[aps,showpacs,twocolumn]{revtex4}

\usepackage{graphicx}
\usepackage{epsfig}
\usepackage{epstopdf}
\usepackage{amsfonts}
\usepackage{amsmath,amssymb}

\begin{document}

\title{Exact Solution of the Isovector Proton Neutron Pairing Hamiltonian}
\author{J.~Dukelsky$^1$,
V.G.~Gueorguiev$^{2,3}$, P.~Van Isacker$^4$, S.~Dimitrova$^3$,
B.~Errea$^1$ and S.~Lerma H.$^1$}
\address{
$^1$Instituto de Estructura de la Materia, CSIC. Serrano 123, 28006
Madrid, Spain \\
$^2$Lawrence Livermore National Laboratory, Livermore, California,
USA\\
$^3$Institute of Nuclear Research
and Nuclear Energy, BAS, Sofia 1784, Bulgaria\\
$^4$Grand Acc\'el\'erateur National d'Ions Lourds, BP 55027,
F-14076 Caen Cedex 5, France}

\date{\today}

\begin{abstract}
The complete exact solution
of the $T=1$ neutron-proton pairing Hamiltonian
is presented
in the context of the SO(5) Richardson-Gaudin model
with non-degenerate single-particle levels
and including isospin-symmetry breaking terms.
The power of the method is illustrated
with a numerical calculation for  $^{64}$Ge
for a $pf+g_{9/2}$ model space
which is out of reach of modern shell-model codes.
\end{abstract}

\maketitle

Exactly solvable models (ESM) % built upon a dynamical symmetry
provide important insights into the structure of many-body quantum
systems. The two main advantages of ESMs are: (1) They can
describe in an analytical or exact numerical way a wide variety of
elementary phenomena. (2) They can be and have been used as a
testing ground for various many-body approaches.

A particular class of ESMs, extensively used in nuclear physics,
are the dynamical-symmetry models. In this case the Hamiltonian
can be expressed in terms of Casimir operators of a chain of
nested algebras. An example often used to introduce nuclear
superconductivity (see {\it e.g.} Ref.~\cite{Rin80}) is the rank-1
(Lie) algebra SU(2). Examples of dynamical-symmetry models
associated with a rank-2 algebra are Elliott's SU(3) model of
nuclear deformation~\cite{Ell58} and the SO(5) model of $T=1$
isovector pairing between neutrons and protons~\cite{Flo52} which
has found many applications in nuclei (see {\it e.g.}
Ref.~\cite{so5}).

% Elliott's Hamiltonian is a linear
%combination of the quadratic Casimir operators of SU(3) (involving
%a quadrupole interaction) and of its SO(3) angular-momentum
%subalgebra. Lie algebras with higher rank lead to more complex
%ESMs: Wigner's SU(4) supermultiplet model~\cite{Wig37}, the SO(5)
%model of $T=1$ isovector pairing~\cite{Flo52}, the SO(8) model of
%$T=0$ isoscalar and $T=1$ isovector pairing~\cite{Flo64,Eva81},
%the extension of SU(3) to the symplectic Sp(6) model~\cite{Ros77},
%and Ginocchio's SO(8) model~\cite{Gin80}, also known as the
%fermion dynamical symmetry model (FDSM)~\cite{Wu87}. The three
%dynamical symmetries of the interacting boson model~\cite{Iac87}
%provide another example.

The concept of quantum integrability, closely linked with exact
solvability, goes beyond the limits of the dynamical-symmetry
approach. A quantum system is integrable if there exist as many
commuting Hermitian operators (integrals of motion) as quantum
degrees of freedom~\cite{Zha95}. The set of Casimir operators of a
chain of nested algebras satisfies this condition.

Dynamical-symmetry models are usually defined for degenerate
single-particle levels. Lifting this degeneracy breaks the
dynamical symmetry but may still preserve integrability. The
pairing model with non-degenerate single-particle levels, of which
an exact solution was found by Richardson in the
sixties~\cite{Ric63}, represents an example of an ESM with such
characteristics. Recently, more general exactly solvable pairing
models, both for fermions and for bosons, called Richardson-Gaudin
(RG) models, have been proposed~\cite{Duk01,Duk04}.

The RG pairing models are based on rank-1 algebras: SU(2) for
fermions and SU(1,1) for bosons. In this Letter we carry out the
first step in extending the RG models to higher-rank algebras by
considering a RG model based on the rank-2 algebra SO(5). The
model Hamiltonian describes a two-component system consisting of
neutrons and protons interacting through an isovector ($T=1$)
pairing force and distributed over non-degenerate orbits. This
neutron-proton (np) pairing Hamiltonian with non-degenerate orbits
has been studied by Richardson~\cite{Ric66} who proposed an exact
solution. However, it was shown subsequently that Richardson's
solution is incorrect for more than two nucleon pairs~\cite{Pan02}
by explicitly solving the case of three-nucleon pairs.
Independently, Links {\it et al.} derived an exact solution for
the isospin invariant SO(5) model by making use of the quantum
inverse scattering method~\cite{Lin02}.

We present here the most general exact solution of the RG SO(5) model
including isospin symmetry breaking terms and for states with arbitray seniority.
In addition to the construction of the complete set of integrals of motion
from which more general exactly solvable pn-pairing Hamiltonians can be derived,
we present here the first numerical exact solution of the SO(5) RG model for $^{64}$Ge
in a Hilbert space built from the $pf+g_{9/2}$ shells,
of which the dimension
goes well beyond the limits of modern shell-model codes
based on exact diagonalization.

SO(5) has also been proposed as the symmetry underlying
high-$T_{\rm c}$ superconductivity~\cite{Dem04}. The exactly
solvable RG model discussed in this Letter may conceivably be used
to generalize SO(5) condensed-matter models~\cite{Wu03} by the
explicit {\em addition} of non-degenerate single-particle
symmetry-breaking terms. Other possible applications might be
found in polarized ultracold Fermi gases with $p$-wave pairing
interactions \cite{Gu}.

We begin by introducing the 10 generators of the SO(5) algebra in
a representation well suited for nuclear physics problems. Let us
define first the three $T=1$ pair-creation operators: $\hat
b^\dag_{-1,i}=\hat n_i^\dag\hat n_{\bar\imath}^\dag$, $\hat
b^\dag_{0,i}=(\hat n_i^\dag\hat p_{\bar\imath}^\dag+\hat
p_i^\dag\hat n_{\bar\imath}^\dag)/\sqrt{2}$, and $\hat
b^\dag_{+1,i}=\hat p_i^\dag\hat p_{\bar\imath}^\dag$, where $n$
and $p$ refer to neutrons and protons, respectively, and $i$
labels a single-particle basis (with ${\bar\imath}$ its
time-reversed state) which may be associated with the spherical
shell-model basis $i\equiv jm$ or with an axially-symmetric
deformed basis $i\equiv\alpha m$. The three pair-annihilation
operators are $\hat b_{-1,i}$, $\hat b_{0,i}$, and $\hat
b_{+1,i}$. The three components of the isospin operator [$\hat
T_{+,i}=(\hat p_i^\dag\hat n_i+\hat p_{\bar\imath}^\dag\hat
n_{\bar\imath})/\sqrt{2}$, $\hat T_{0,i}=(\hat p_i^\dag\hat
p_i+\hat p_{\bar\imath}^\dag\hat p_{\bar\imath})/2 -(\hat
n_i^\dag\hat n_i+\hat n_{\bar\imath}^\dag\hat n_{\bar\imath})/2$,
and $\hat T_{-,i} =(\hat n_i^\dag\hat p_i+\hat
n_{\bar\imath}^\dag\hat p_{\bar\imath})/\sqrt{2}$] close the ${\rm
SU}_T(2)$ subalgebra of SO(5). These 9 operators together with the
number operator $\hat N_{i}=\hat p_i^\dag\hat p_i+\hat
p_{\bar\imath}^\dag\hat p_{\bar\imath}+ \hat n_i^\dag\hat n_i+\hat
n_{\bar\imath}^\dag\hat n_{\bar\imath}$ define the SO(5) algebra.

For a system with $L$ single-particle states $i=1,...,L$
there are $L$ integrals of motion:
\begin{eqnarray}
\hat R_{i} &=&
(2+\Delta)\hat H_i+\Delta\hat T_{0,i}
+2g\sum_{i^\prime(\neq i)=1}^L
\frac{1}{z_{i}-z_{i^\prime}}
\label{Integ}
\\&\times&
\left[\sum_\mu\left(\hat b_{\mu,i}^\dag\hat b_{\mu,i^\prime}+
\hat b_{\mu,i^\prime}^\dag\hat b_{\mu,i}\right)
+\hat T_i\cdot\hat T_{i^\prime}+\hat H_i\hat H_{i^\prime}\right],
\nonumber
\end{eqnarray}
where $\hat H_i=\hat N_i /2 -1$.
The expression~(\ref{Integ}) follows from the integrals of motion
valid for any semi-simple algebra of arbitrary rank~\cite{Aso02}.
Since SO(5) is of rank 2,
its Cartan subalgebra contains two elements
namely $\hat H_i$ and $\hat T_{0,i}$ in the chosen basis
which therefore appear linearly in~(\ref{Integ}).
The set of $L$ parameters $z_i$
together with the two constants $g$ and $\Delta$
can be freely chosen
and it is straightforward to check
that the integrability condition $[\hat R_i,\hat R_j]=0$ is valid
for any choice of the $L+2$ parameters.
A simplified version of~(\ref{Integ}) was previously derived
using the algebraic Bethe {\it ansatz}~\cite{Lin02}.

The eigenvalues of the integrals of motion are
\begin{eqnarray}
r_i&=&
2g\!\!\!\!\!\!\sum_{i^\prime(\neq i)=1}^L
\frac{
\left(\frac{v_i}{2}-1\right)
\left(\frac{v_{i^\prime}}{2}-1\right)+
t_it_{i^\prime}}{z_i-z_{i^\prime}}-
g\!\!\!\!\!\!\sum_{\beta=1}^{M+T_0+t}
\frac{2t_i}{z_i-\omega_\beta}
\nonumber\\
&+&g\sum_{\alpha=1}^M
\frac{v_i+2t_i-2}{z_i-e_\alpha}+
\left[(2+\Delta)\left(\frac{v_i}{2}-1\right)-\Delta t_i\right]\!,
\label{Eigr}
\end{eqnarray}
where $e_\alpha$ and $\omega_\beta$
are solutions of the equations
\begin{eqnarray}
\frac{1}{g}&=&
\frac{1}{2}\sum_{i=1}^L
\frac{v_{i}+2t_i-2}{z_i-e_\alpha}+
\sum_{\alpha^\prime(\neq\alpha)=1}^M
\frac{2}{e_{\alpha^\prime}-e_\alpha}
\nonumber\\
&-&\sum_{\beta=1}^{M+T_0+t}
\frac{1}{\omega_\beta-e_\alpha},
\label{Richeqs}\\
\frac{\Delta }{g}&=&
\sum_{\beta^\prime(\neq\beta)=1}^{M+T_0+t}
\frac{2}{\omega_{\beta^\prime}-\omega_\beta}-
\sum_{\alpha=1}^M
\frac{2}{e_\alpha-\omega_\beta}-
\sum_{i=1}^L\frac{2t_i}{z_i-\omega_\beta}.
\nonumber
\end{eqnarray}
The meaning of the  quantum numbers appearing in~(\ref{Eigr})
and~(\ref{Richeqs}) is as follows: $v_i$ is the seniority of each
$i$ level, {\it i.e.} the number of fermions not paired in
time-reversed states with isospin $T=1$, $t_i$ is the isospin of
the unpaired fermions (this quantum number is often called reduced
isospin~\cite{Hec65}), $t=\sum_it_i$, $M$ is the number of $T=1$
time-reversed pairs, and $T_0$ is the $z$ component of the total
isospin, {\it i.e.} the eigenvalue of the operator $\hat
T_0=\sum_i\hat T_{0,i}$. The total number of nucleons is
$N=N_p+N_n=2M+\sum_i v_i$ whereas their difference is
$N_p-N_n=2T_0$. The quantum numbers $M$, $T_0$, $v_i$, and $t_i$
are conserved; $T$ is also conserved if $\Delta=0$.

Although any function of the $\hat R_i$
can be used as an integrable Hamiltonian,
the linear combination $\sum_{i=1}^L z_i\hat R_i$
yields simple expressions for the np-pairing Hamiltonian
and its corresponding eigenvalues:
\begin{eqnarray}
\hat H&=& \frac{1}{2} \sum_{i=1}^L  z_i\hat R_i + g \hat C
\label{Ham1}\\
&=&\sum_j\varepsilon_j
\left(\hat N_j+\Delta\hat N_{p,j}\right)
+\frac{g}{2}\hat T\cdot\hat T+
g\!\!\!\!\sum_{\mu jmj^\prime m^\prime}\!\!\!
\hat b_{\mu,jm}^\dag\hat b_{\mu,j^{\prime }m^{\prime }},
\nonumber
\end{eqnarray}
where $\hat C$ is a constant operator depending on the conserved
quantities.  We have introduced the variables
$\varepsilon_j=z_j/2$ and specialized to a spherical basis $i
\equiv jm$. The second term on the r.h.s. of~(\ref{Integ}) breaks
the isospin symmetry. For $\Delta\ne0$ the operator $\hat T^2$
does not commute with the Hamiltonian~(\ref{Ham1}) and,
consequently, $T$ is not a good quantum number but $T_0$ is still
a conserved quantity. The  same linear combination of the $r_i$
and the use of the Richardson equations~(\ref{Richeqs}), yield the
eigenvalues of~(\ref{Ham1}):
\begin{eqnarray}
E&=&\sum_{\alpha=1}^M e_\alpha+
\frac{\Delta}{2}\!\sum_{\beta=1}^{M+T_0+t}\!w_\beta+
\sum_j\varepsilon_j \left[\frac{v_j}{2}(2+\Delta)-\Delta
t_j\right] \nonumber\\&&+ \frac{g}{2} T_0(T_0-1).\label{ener}
\end{eqnarray}

\begin{figure}
%\vspace{-0.3cm} \hspace{-.8cm}
\includegraphics[width=7.3cm]{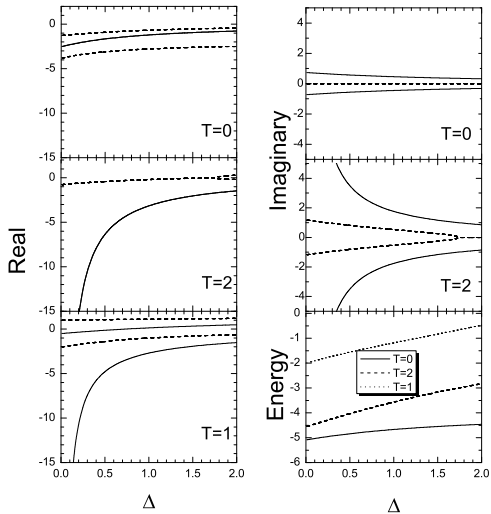}
\caption{Pair energies $e_\alpha$ (dashed lines) and spectral
parameters $\omega_\beta$ (solid lines) as a function of the
parameter $\Delta$ for three different states of a 2n-2p system
with two shells ($j_0=1/2$ and $j_1=3/2$) in the seniority-0
subspace. The label $T$ refers to the isospin in the limit
$\Delta=0$. The figure shows the lowest-energy states of the
Hamiltonian~(\ref{Ham1}) for each $T$ with energies plotted in the
bottom-right panel. Since the spectral parameters of the $T=1$
state are real, their imaginary part is not shown. The interaction
strength and single-particle energies are $g=-1$,
$\varepsilon_0=0$, and $\varepsilon_1=1$.} \label{EvsDel}
\end{figure}

Each solution of the equations~(\ref{Richeqs})
gives an eigenstate of the np-pairing Hamiltonian.
The spectral parameters $e_\alpha$
are interpreted as pair energies as in the case of SU(2) pairing.
%as suggested by the eigenvalues of the p-n pairing Hamiltonian.
However, due to the larger rank of SO(5), a new set of spectral
parameters $w_\beta$ appears in the equations~(\ref{Richeqs}).
These new parameters are associated with the ${\rm SU}_T(2)$
isospin subalgebra and there are $M+T_0+t$ of them. In the limit
$\Delta=0$ the number of finite $w_\beta$ parameters reduces to
$M+t-T$ for each possible isospin $T$. The Bethe {\it ansatz} for
the SO(5) eigenstates of the RG model is a product wave
function~\cite{Ush94}:

\begin{equation}
%\vert\Psi\rangle=
\left(\prod\limits_{\alpha=1}^M\hat b_{-1}^\dag(e_\alpha )\right)
\prod\limits_{\beta=1}^{M+T_0+t}
\left(\hat T_+(w_\beta)-\sum_{\alpha=1}^M\frac{\hat I_{+,\alpha}}{w_\beta-e_\alpha}\right)
\vert\Lambda\rangle,
\label{Ansa}
\end{equation}
where the spectral dependence of the operators is
\begin{equation}
\hat T_+(w_\beta)=\
\sum_i \frac{\hat T_{+,i}}{2\varepsilon_i-w_\beta},
\quad
\hat b_\mu^\dag(e_\alpha)=
\sum_i\frac{\hat b^\dag_{\mu,i}}{2\varepsilon_i-e_\alpha},
\label{Oper}
\end{equation}
and $\hat I_{+,\alpha}$ is a raising  operator for $\hat
b_\mu^\dag(e_\alpha)$: $\hat b_\mu^\dag(e_\alpha)\hat
I_{+,\alpha}= \hat b_{\mu+1}^\dag(e_\alpha)$ for $\mu=-1,0$, $\hat
b_{+1}^\dag(e_\alpha)\hat I_{+,\alpha}=0$, and
$\vert\Lambda\rangle$ is a lowest-weight state defined by $\hat
b_{\mu,i}\vert\Lambda\rangle=\hat T_{-,i}\vert\Lambda\rangle=0$.
To show the behavior of the spectral parameters $e_\alpha$ and
$\omega_\beta$ as a function of the isospin-breaking term
$\Delta$, we plot in Fig.~\ref{EvsDel} some selected solutions of
the Richardson equations~(\ref{Richeqs}) for a system of two
neutrons and two protons in two shells ($j_0=1/2$ and $j_1=3/2$)
within the seniority-0 subspace ($M=2,T_0=0,t_i=v_i=0$).

For $\Delta=0$
%the solutions are labeled by the total isospin $T$.
there are two finite $\omega_\beta$ complex conjugate parameters
for $T=0$ while the two pair energies are real and negative, for
$T=1$ there is one real, finite $\omega_\beta$ and two real pair
energies $e_\alpha$, and, finally, the $T=2$ case reduces to SU(2)
for like particles with two $e_\alpha$ parameters forming a
complex conjugate pair. For finite $\Delta$ there is isospin
mixing and the number of finite parameters $\omega_\beta $ is
always two. Figure~\ref{EvsDel} thus confirms that the number of
finite $\omega_\beta$ spectral parameters reduces from $M=2$ to
$M-T$ when $\Delta \rightarrow 0$. For $T=1$ one of the real
$\omega_\beta$ goes to $-\infty$ in this limit, vanishing from the
Richardson equations (\ref{Richeqs}) but giving a finite
contribution to the Hamiltonian eigenvalues ({\ref{Ham1})
\cite{note}. Analogously, in the $T=2$ case the two $\omega$
parameters tend to $\infty$ in the $\Delta=0$ limit. Also shown
are the energies of the three eigenstates of the np-pairing
Hamiltonian~(\ref{Ham1}). We emphasize that, while these are the
eigenvalues of a particular Hamiltonian, the spectral parameters
completely define the eigenfunction~(\ref{Ansa}) of the $L$
integrals of motion~(\ref{Integ}) from which $\hat H$ is
constructed and their corresponding eigenvalues (\ref{Eigr}).

We now turn to the discussion of a numerical calculation for
$^{64}$Ge. We consider a model space that is well beyond modern
shell-model capabilities based on exact diagonalization: 12
valence neutrons and 12 valence protons with a $^{40}$Ca core. The
adopted single-particle energies are (in MeV)
$\varepsilon_{f_{7/2}}=0.00$, $\varepsilon_{p_{3/2}}=6.00$,
$\varepsilon_{f_{5/2}}=6.25$, $\varepsilon_{p_{1/2}}=7.1$, and
$\varepsilon_{g_{9/2}}=9.60$, and two pairing strengths, $g=-0.05$
(weak) and $-0.5$ (strong), are considered. We assume isospin
symmetry ($\Delta=0$) and consider the seniority-0 subspace.

Results for the lowest $T=0$, 1, and 2 states
are shown in Fig.~\ref{T2}.
\begin{figure}
%\vspace{-0.3cm} \hspace{-.8cm}
\includegraphics[width=7.3cm]{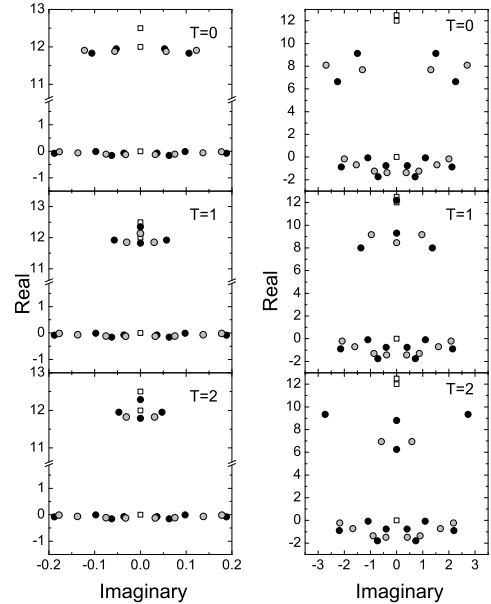}
\caption{Complex-plane representation of the pair energies
$e_\alpha$ and spectral parameters $\omega_\beta$ for the
lowest-energy states with isospin $T=0,1,2$ in $^{64}$Ge. The left
panel corresponds to weak coupling $g=-0.05$ and the right panel
to strong coupling $g=-0.5$. The  squares represent the three
lowest single-particle energies ($2\varepsilon_{f_{7/2}}=0.00$,
$2\varepsilon_{p_{3/2}}=12$, $2\varepsilon_{f_{5/2}}=12.5$), the
black circles are the pair energies $e_\alpha$, and the grey
circles are the parameters $\omega_\beta$. All energies are in
MeV.} \label{T2}
\end{figure}
The $T=0$ solution corresponds to the ground state while the $T=1$
and $T=2$ solutions are excited states in $^{64}$Ge. As in SU(2),
the different configurations can be classified in the
weak-coupling limit. At weak coupling ($g \rightarrow 0$) eight
pairs occupy the $f_{7/2}$ level and four pairs are in the
$p_{3/2}$ level for the state with $T=0$. This is reflected in the
upper left panel of figure \ref{T2} where 8 pair energies appear
close to $2 \varepsilon_{f_{7/2}}$ and 4 pair energies are close
to $2 \varepsilon_{p_{3/2}}$ making the corresponding terms in
(\ref{Oper}) dominant. Due to the Pauli principle, this
configuration is not allowed for a state with $T=1$ and,
correspondingly, one pair energy is close to
$2\varepsilon_{f_{5/2}}$. In all cases the $w_\beta$ parameters
are intertwined with the pair energies $e_\alpha$. The number of
$w_\beta$ parameters ($M-T$), together with the initial
configuration at weak coupling, defines each eigenstate of the
np-pairing Hamiltonian. As $|g|$ increases, the $e_\alpha$ and
$w_\beta$ parameters expand in the complex plain. The solutions
are subject to numerical instabilities due to singularities
arising when a real pair energy $e_\alpha$ crosses a
single-particle energy or when real $e_\alpha$ and $w_\beta$
parameters cross. An example of the first class of crossings can
be observed in Fig.~\ref{T2} for $T=2$ where the pair energy above
the $p_{3/2}$ level at weak coupling goes down with increasing $g$
and crosses the $p_{3/2}$ single-particle energy. The $T=1$ case
shows an exchange of positions on the real axis of a pair energy
$e_\alpha$ and a $w_\beta$ parameter as an example of the second
class of singularities. The first class of singularities was
already present in SU(2) pairing and precluded the practical use
of Richardson's solution for a long time. Recently, a new method
to overcome this numerical problem was proposed~\cite{Rom04}. We
believe that the same procedure can be used to treat the second
class of singularities as well, allowing the exact solution of the
SO(5) model for very large systems.

As a further illustration of the method we show in Fig.~\ref{T3}
the eigenenergies and the occupation probablities of single-particle levels
as a function the pairing strength.
\begin{figure}
%\vspace{-0.3cm} \hspace{-.8cm}
\includegraphics[width=7.3cm]{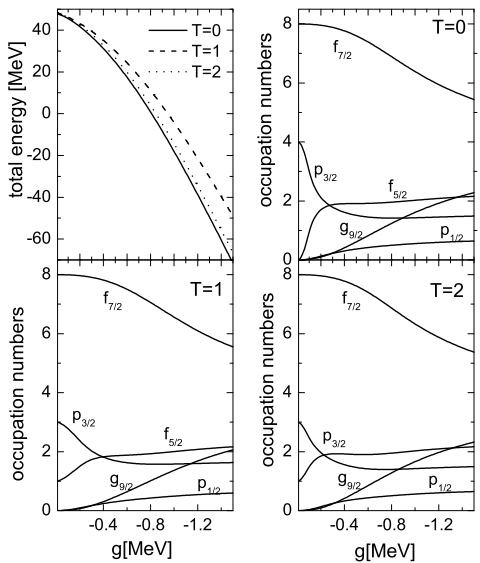}
\caption{Eigenenergies and occupation probablities of single-particle levels
of the $T=0$, 1, and 2 states in $^{64}$Ge
as a function of the pairing strength $g$.}
\label{T3}
\end{figure}
The occupation probabilities can be obtained making use of
Hellmann-Feynman theorem which expresses them in terms of
derivatives of the eigenvalues of the integrals of motion $r_i$ as
$\langle N_{p,i}\rangle= 1+\frac{\partial r_i}{\partial\Delta}$,
and $ \langle N_{n,i}\rangle= r_i-g\frac{\partial r_i}{\partial
g}- (1+\Delta )\left(1+\frac{\partial
r_i}{\partial\Delta}\right)+1$. These derivatives can be related
to the derivatives of the spectral parameters $e_\alpha$ and
$\omega_{\alpha}$, which in turn can be obtained taking the
derivatives of the Richardson equations~(\ref{Richeqs}).

In summary, as an application of generalized RG models, we have
presented the complete solution of the SO(5) isovector np-pairing
problem. The generalization allows the introduction of one-body
symmetry-breaking terms, such as non-degenerate single-particle
energies, yielding an exact solution of the SO(5) np-pairing model
for arbitrary seniorities even if it includes an isospin-breaking
term. The numerical solution of the SO(5) Richardson equations was
presented for the specific example of $^{64}$Ge, together with a
discussion of the behavior of the spectral parameters for weak and
strong pairing. With this work the exact solution for large
systems with SO(5) symmetry is now available which could be of
great importance in condensed-matter physics when addressing the
phenomenon of high-$T_{\rm c}$
superconductivity~\cite{Dem04,Wu03}. Finally, the treatment of
higher-rank algebras like Sp(6) and SO(8) opens the possibility of
exact nuclear structure calculations with more realistic quantum
integrable models.

This work was supported in part by the Spanish DGI under grant
No.~BFM2003-05316-C02-02, by Bulgarian contracts $\Phi-1416$ and
$\Phi-1501$ and by a CICYT-IN2P3 cooperation. V.G.G. acknowledges
financial support from a NATO fellowship and from the US DOE under
grant No.~W-7405-Eng-48. S.L.H. has a post-doctoral fellowship
from the Spanish SEUI-MEC and B.E. has a pre-doctoral grant from
the Spanish CE-CAM.


\begin{thebibliography}{00}
\bibitem{note} It can be shown that in the limit $\Delta \rightarrow 0$
the second term for the diverging $\omega's$ and the fourth term
in (\ref{ener}) combine to give $\frac{g}{2} T (T+1)$.
\end{thebibliography}
\end{document}